# CS EVAPORATION IN A NEGATIVE ION SOURCE AND CS CLEANING TESTS BY PLASMA SPUTTERING*

M. Barbisan[1,ξ], R.S. Delogu[1], A. Pimazzoni[1], C. Poggi[1,2], M. Ugoletti[1,2] and M. Cavenago[3]

[1]*Consorzio RFX (CNR, ENEA, INFN, Università di Padova, Acciaierie Venete SpA), C.so Stati Uniti 4, I -35127, Padova, Italy*
[2]*Università degli Studi di Padova, via 8 Febbraio 2, I-35122, Padova, Italy*
[3]*INFN-LNL, v.le dell'Università n. 2, I-35020, Legnaro (PD) Italy*

*Abstract*

The compact radio frequency negative ion source NIO1 (Negative Ion Optimization phase 1) has been designed, built and operated by Consorzio RFX and INFN-LNL in order to study and optimize the production and acceleration of H$^-$ ions in continuous operation. In 2020 Cs was evaporated in the source to increase the total extracted ion current. After an initial reduction of extracted electron to ion ratio and subsequently an increase of extracted negative ion current, the source performances progressively worsened, because of the excessive amount of Cs evaporated in the source; the extracted electron to ion ratio increased from below 1 to more than 10, while ion current density reduced from max. 67 A/m$^2$ ion current to not more than 30 A/m$^2$). The paper presents the experimental observations collected during Cs evaporation (reduction of plasma light, Cs emission and H$_\beta$/H$_\gamma$ ratio, etc.) that can help stopping the process before an excessive amount of Cs is introduced in the source. The paper also reports the cleaning techniques tested to remove the Cs excess by the action of hydrogen or argon plasmas; while argon was predictably more effective in surface sputtering, a 3 h Ar plasma treatment was not sufficient to recover from overcesiation.

## I. INTRODUCTION

NIO1 (Negative Ion Optimization phase 1) is a radiofrequency H$^-$ ion source, built and operated by Consorzio RFX and INFN-LNL, with the aim of studying the production and acceleration of negative ions, to address present and future issues and needs of negative ions sources for neutral beam injectors in fusion reactor (DEMO in particular) [[1]-[4]]. The NIO1 source is designed to be compact and flexible for the test of new concepts, components and instrumentation. It can be operated for several hours with plasma and beam extraction, giving the possibility to study physics and operating scenarios that are not accessible to the larger ion sources for fusion application up to now [[5]-[7]]. As shown in Figure 1, the source volume is a 21 cm x Ø10 cm cylinder; the plasma is sustained by 2 MHz, max. 2.5 kW radiofrequency (RF), radiated from a solenoid through a pyrex cylinder. The other inner surfaces (excluding the acceleration system) are protected by a molybdenum foil; multipole magnet systems reduce the plasma-surface interaction. The source needs to be operated at gas pressure levels that are higher than the 0.3 Pa target level for ITER (and likely DEMO) [[8]-[10]], because the mean free path of electrons must be kept within the dimensions of the source, which are smaller than other ITER/DEMO relevant negative ion sources [[8]-[10]].

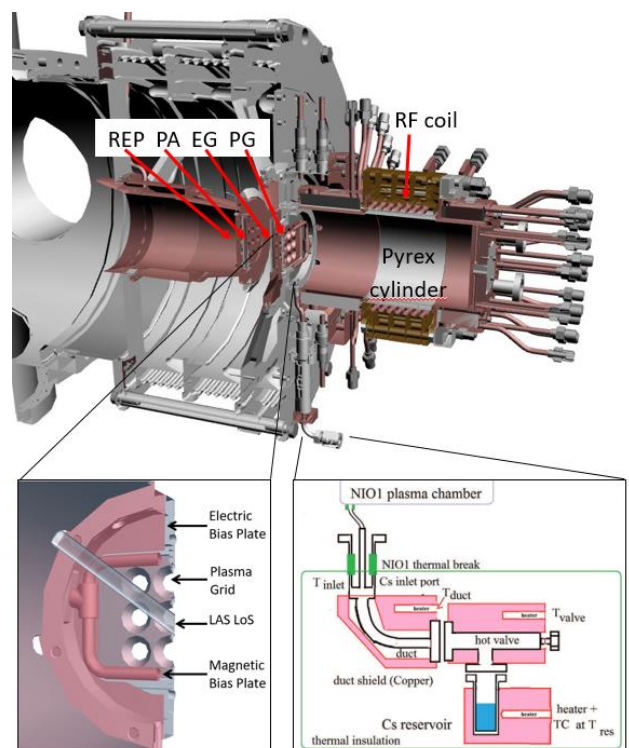

**Figure 1.** a) Vertical section of NIO1 3D model, with a detailed representation of the bias plates zone (bottom left) and of the Cs oven structure (bottom right).

* This work has been carried out within the framework of the EUROfusion Consortium and has received funding from the Euratom research and training programme 2014-2018 and 2019-2020 under grant agreement No 633053. The views and opinions expressed herein do not necessarily reflect those of the European Commission.

ξ email: marco.barbisan@igi.cnr.it

The negative ions are extracted through a 4-grid system (plasma grid – PG, extraction grid – EG, post acceleration grid – PA, repeller grid – REP); co-extracted electrons are mostly dumped to the EG thanks to the magnetic field generated by its embedded magnets. Keeping the fraction of co-extracted electrons as low as possible is a fundamental target for the sources of future neutral beam injectors, and resulted to be challenging for long plasma operation in other experiments [11]. The diffusion of electrons towards the PG is limited by a vertical magnetic field [12] (from 9 mT to 12.5 mT in proximity of the PG, as configuration F3 in reference [4]). The magnetic field is generated by permanent magnets and by a current flowing in the PG. For the same purpose, positive potential differences can be established [[12]-[16]] between the PG and the source body, and between the two bias plates (Figure 1, bottom left view), electrodes placed just upstream the PG (see figs. 2 and 10 of ref.[1]), and the source body. During the experimental campaign, which is object of this paper, the two bias plates were electrically connected outside vacuum and will be hereafter referred as a single bias plate (BP). The biases are such to attract and remove electrons at the PG, moreover they modify the potential drop of the plasma sheath at the PG [[12]-[16]].

Negative ions are generated in the plasma volume through dissociative electron attachment reactions, involving ro-vibrationally excited hydrogen molecules. A much higher amount of negative ions can be made available at the PG apertures thanks to surface reactions on the PG, that convert neutrals and positive ions into $H^-$ [[12],[17]-[19]]. The rate of the surface reactions are exponentially dependent on the surface work function, in turn dependent on the thickness and purity of the Cs layer [18]. The increase of negative ions at the PG, made possible by Cs evaporation, is not just beneficial in increasing the extracted ion current. Because of plasma quasi-neutrality, the higher amount of $H^-$ at the PG locally reduces the density of electrons that could then be co-extracted.

Cs evaporation in NIO1 is performed by means of an oven (Figure 1, bottom right view), composed of several sections with independent heating and PID temperature control. There are a Cs reservoir, a manual valve to protect the reservoir from air or impurities during holydays or maintenance activities, and the first section of the duct which leads the Cs vapour towards the source volume. These oven components are insulated in glass wool; the duct, with 4 mm inner diameter and 0.2 m length, reaches the source inside a larger vacuum tight tube, and connects to the bias plate by means of a macor insert. In this way, the duct can be kept at higher temperature than the bias plate, to reduce the risk of Cs accumulation on cold spots.

The present paper will show the results of the first experimental campaign in NIO1 with Cs evaporation, describing the effects of Cs on the source performances, from the first detection of Cs up to a condition of over-cesiation. The paper will also report the results of several non-invasive tests performed with the aim of cleaning the source from the excess of Cs. The plasma conditions were mostly monitored by three diagnostics:

- Plasma light detection: a photomultipler (PMT), active in the 230 nm÷920 nm wavelength range, measures the light intensity collected by an optic head placed on the rear side of the source.
- Optical Emission Spectroscopy (OES) [20]: a survey spectrometer (0.3 nm/pix.) and a high resolution spectrometer (7 pm/pix.) monitor the intensity of hydrogen and Cs spectral lines between 200 nm and 900 nm; both systems are wavelength and intensity calibrated. The light is collected by a common optic head on the rear side of the source.
- Laser Absorption Spectroscopy (LAS): a laser beam crosses the source volume at 19 mm distance from the PG (Figure 1, bottom left view), scanning its wavelength around the 852 nm Cs D2 line; from the measurement of the absorption spectrum it is possible to estimate the line-averaged density of ground-state neutral cesium density. The setup is similar to the LAS diagnostics installed in several other negative ion sources [[21]-[23]]; the laser intensity is low enough to avoid ground state depopulation effects that would lead to an underestimation of measurements [[22]-[23]].

A Surface Ionization Detector (SID), while being desirable to measure the Cs flux from the oven duct [[24]-[25]], could not be installed due to space limitations. Numerical simulations are planned in future to estimate the Cs flux from the temperatures of the oven components.

Even if the outcomes of the cleaning tests were not fully successful, the authors believe that this contribution will be helpful to avoid over-caesiation of negative ion sources and to the investigation of means for cleaning a caesiated source while avoiding Cs interaction with air. This issue will become particularly relevant in the case of the $H^-/D^-$ sources in ITER neutral beam injectors, whose accessibility is very limited and which may even be contaminated by tritium.

## II. EXPERIMENTAL RESULTS

### A. Cs evaporation and over-cesiation

The NIO1 source was preliminarily cleaned by simple plasma action, until no trace of oxygen or OH emission was detectable through OES. It is known from other experiments that the source body temperature has to be raised to about 40 °C to guarantee a proper Cs redistribution, while the PG temperature must be even higher to minimize its work function [[13],[26]]. In NIO1, for this first campaign, the cooling system was set at the maximum temperature allowed by the water cooled devices, so the source and the PG were kept between 30°C and 40°C. Attempts to successfully evaporate Cs into the source were performed with increasing oven temperatures. Cs was first detected on 1st Sept. 2020, with reservoir temperature $T_{res}$ and valve temperature $T_{valve}$ set at 200°C, while the duct temperature was set at $T_{duct}$=250°C. The operative conditions were: input RF power $P_{RF}$=1500 W (the maximum temporarily allowed for the source safety was 1700 W), source pressure $p_{source}$=0.75 Pa, PG filter current $I_{PG}$=400A (the maximum allowed by the power supply, for maximum electron filtering); negative ions were extracted with $U_{ex}$=0.7 kV extraction voltage (PG-EG) and $U_{tot}$=6 kV total acceleration voltage (PG-PA). The action of Cs was evident from a clear decrease of the EG current, which in volume operation is supposed to be mainly composed of co-extracted electrons, from 50 mA to 14 mA in 3.5 h of continuous plasma and beam extraction. This clearly indicated a modification of the e/H$^-$ density ratio at the PG surface, despite during this time the extracted beam did not substantially increase in current density (about 15 A/m$^2$). With a fixed PG-source body bias of 10 V, a strong increase of the bias current was observed, from 0.16 A to 0.6 A (the limit of the power supply, subsequently substituted). The BP-source body bias (13 V, 0.8 A) did not show relevant variations during cesiation process. This can be interpreted as an increase of the plasma potential drop at the PG, due to the local higher presence of negative ions. The emission lines of Cs at 672 nm, 852 nm and 894 nm were clearly detectable by OES. Cs evaporation, plasma and beam extraction, which lasted 5 h., were resumed on the following day for 3 h., operating the oven from $T_{res}$=200°C to $T_{res}$=240°C ($T_{duct}$ from 250°C to 260°C). With these values of oven temperatures, the highest value of extracted beam current density in NIO1 was measured: 67 A/m$^2$, with $U_{ex}$=1.2 kV, $U_{tot}$=11 kV, $P_{RF}$=1600 W, $P_{source}$=0.75 Pa, $I_{PG}$=10 A (the filtering by the permanent magnets was enough, thanks to the already lower electron current); PG-source body bias was set at 15V (current ~1.5A, while the BP-source body bias was set at 0.3 A (voltage ~9.5 V). In these conditions, the extracted beam current was roughly equal to the co-extracted electron current. In the four subsequent experimentation days (between 4$^{th}$ and 11$^{th}$ September), Cs evaporation and plasma were performed for about 15 h in total, with $T_{res}$=200°C, $T_{valve}$=220°C and $T_{duct}$=250°C; the source was mostly operated at $P_{source}$=0.6÷0.7 Pa, $P_{RF}$=1500÷1600 W, $I_{PG}$=10 A, PG-source body bias at 15V and BP-source body bias at 0.3 A. A progressive and rapid reduction of beam current density was observed, so that it was not possible to exceed 30 A/m$^2$; the fraction of co-extracted electrons raised from 3 to few tens. The two data likely indicated a reduction of the negative ion density at the PG, due to a higher surface work function and/or to a lower incoming flux of H and H$^+$. This phenomenon was accompanied by a reduction of the plasma luminosity (which is dominated by hydrogen Balmer H$_α$ radiation), as shown in Figure 2, in which the PMT signal is plotted against time over the whole experimental campaign.

More detailed information was provided by the OES diagnostic. Figure 3a shows the emissivity of Balmer H$_β$ and H$_γ$ lines, together with the emissivity of the Cs 2D line, plotted against dataset number (acquisitions of the entire system conditions); dataset ranges with no emissivity data correspond to source conditions with no plasma or unavailability of OES data. The emissivity of H$_β$ and H$_γ$ lines shows an almost regular decrease from 2$^{nd}$ Sept. to 11$^{th}$ Sept.; their ratio (H$_β$/H$_γ$) also progressively decreased, as shown in parallel by Figure 3b. According to a collisional radiative model developed for larger sources, this could be interpreted either as a reduction of electron density or as an increase of electron temperature [27]. In parallel, the Cs D2 line emissivity reached a peak level at 3$^{rd}$ Sept., when the maximum beam current density was reached, but then decreased with time, despite Cs evaporation was constantly active. The most likely explanation is that an excessive amount of Cs was introduced into the source, so that not just the PG optimal Cs covering was exceeded, but that the plasma conditions were also deeply modified, with a lower electron density.

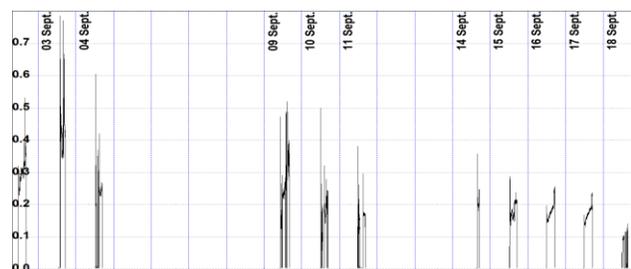

**Figure 2.** Time evolution of the plasma light signal (V) during the experimental campaign.

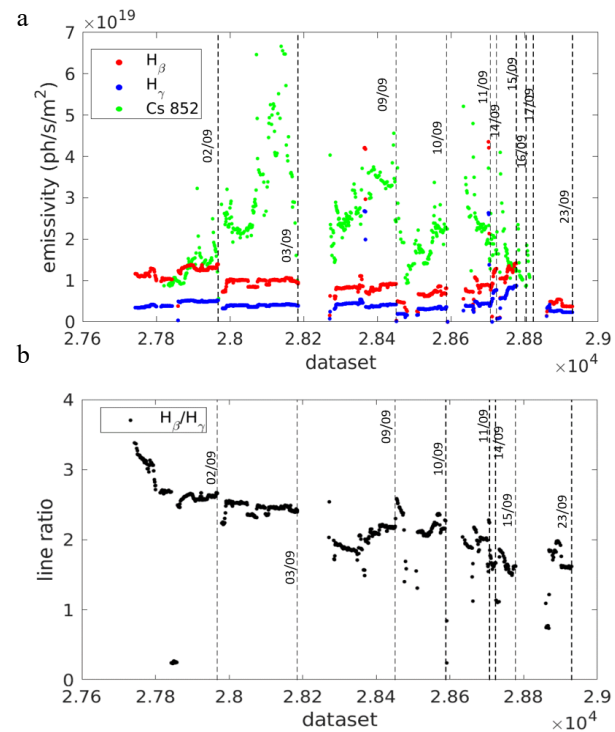

**Figure 3.** H$_β$, H$_γ$, Cs D2 line intensity (a) and H$_β$/H$_γ$ line intensity ratio as function of dataset during the experimental campaign.

### B. Cs density measurements

In the 2-11th Sept. experimental days it was not possible to measure the Cs density by means of the LAS diagnostic: it was not possible to detect the D2 absorption line, despite it was brightly visible in emission spectra. The reason was identified to be in the wavelength range covered by the tunable laser, which was actually shifted from the D2 line wavelength despite the specifications provided by the manufacturer. After having substituted the laser, the first successful absorption spectra were acquired. An example is provided in Figure 4: as the laser current is linearly swept over time (at 1 Hz in the case of NIO1), a corresponding cooling linear variation of the emitted wavelength allows to detect the D2 line in its fine structure: 2 peaks, each one accounting for 3 transitions between states determined by the atom total angular momentum quantum number F [21]. The base of each absorption peak results to be widened by the Zeeman splitting of the atom levels: the LAS line of sight indeed passes through the cylindrical array of magnets close to the PG, where the magnetic filter field can locally reach 100 mT, down to about 10 mT in front of the PG. A similar effect was reported and studied in ref.[28].

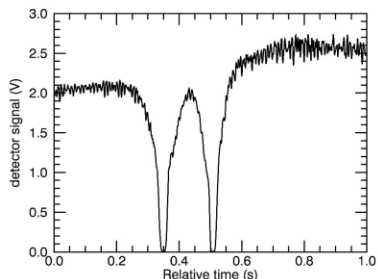

**Figure 4.** Typical LAS absorption spectrum in NIO1.

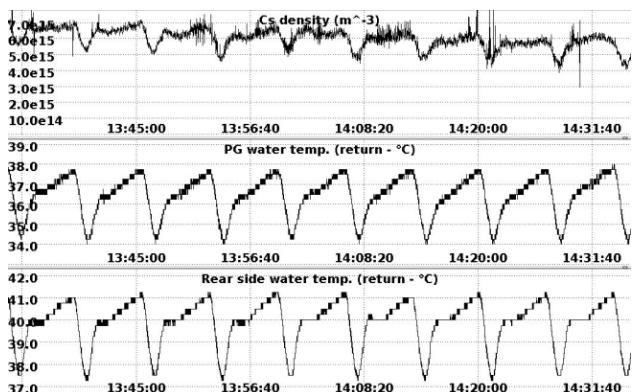

**Figure 5.** Cs density at the PG (plasma phase), water temperature at the cooling return lines of PG and rear plate, as a function of time (16 Sept. 2020, source cleaning tests).

The analysis of the absorption spectra were performed, according to ref. [[21]-[23]] and assuming an effective absorption length of 28.4 cm, i.e. the distance between the vacuum windows from which the laser beam is emitted and collected. The first line-integrated measurements of Cs density were few $10^{16}$ m$^{-3}$ in absence of plasma, one order of magnitude higher with what measured at the IPP negative ion sources BATMAN and ELISE [[29]-[30]]. The high and stable density of Cs, with no plasma and no evaporation, together with the evidence that Cs had deeply drifted through the pipes of the LAS line of sight, add to the evidence that the source was overcesiated.

The LAS diagnostic only measures the density of neutral Cs; comparing the Cs density immediately before and after a switch off of the plasma, and neglecting the redistribution effect of the plasma, the ionization fraction of Cs was estimated to be around 50%÷70%, comparably to measurements performed in another negative ion source [[21],[30],[31]].

The LAS measurements also allowed to show how sensitive Cs density was, in this regime, to the temperature of the source walls. For instance, Figure 5 shows the Cs density and the water temperatures at the return pipes of the PG and source rear block cooling circuits, as a function of time. The oscillation of water temperatures is due to the protection control of the cooling system, which regulates the heat exchange between the NIO1 cooling circuit and the building water cooling system. As shown by the figure, variations of about 4 °C can induce variations of about 20 % in the density of Cs circulating inside the source. Increasing the temperature stability of the cooling system will then be necessary to improve the control of Cs into the source.

### C. Cs cleaning tests

| Test data | No Plasma (Start) | Plasma phase | No Plasma (End) |
|---|---|---|---|
| 15/09/2020 | 20 | Prog. reduction from 16 to 10. | 18 |
| 16/09/2020 | N.A. | Prog. reduction from 9 to 5. | 16 |
| 17/09/2020 | 14 | Prog. reduction from 8 to 3. | 8 |
| 18/09/2020 | 12 | Prog. reduction from 9 to 4. 12 in the no plasma intervals. | 12 |
| 23/09/2020 | 4 | Between 9 and 16 in Ar. Prog. reduction from 3 to 0.8 in hydrogen. | 4 |

**Table 1:** Average values from the LAS diagnostic (in units of $10^{15}$ m$^{-3}$) before, during and after each cleaning test.

In order to remove the excess Cs in the source, without exposing the source to air, the oven was not heated any more; the use of a standard hydrogen plasma was initially considered for cleaning. NIO1 has the possibility to rapidly cumulate plasma-on time thank to the possibility to operate continuously. The following cleaning procedures were tested in separate days:

- Continuous hydrogen plasma and beam extraction (15/09/2020): $P_{RF}$=1.6 kW, $p_{source}$=0.75 Pa, $I_{PG}$=10 A, PG-source body bias 25 V (about 1.6 A), BP-source bias 0.3 A (about 10.5 V). Plasma duration: 4.5 h.

- Continuous hydrogen plasma, no beam extraction and low bias to let $Cs^+$ ions exit the source through the PG apertures (16/09/2020). Beam extraction was temporarily activated at intervals of 30 min. to check beam performances. $P_{RF}$=1.6 kW, $p_{source}$=0.75 Pa, $I_{PG}$=10 A, PG-source body bias 1 V (about 0.6 A), BP-source bias 0.1 A (about 1.0 V). Plasma duration: 5 h.
- Continuous hydrogen plasma, no beam extraction and bias of inverted polarity to make $Cs^+$ ions exit the source through the PG apertures (17/09/2020). Beam extraction was temporarily activated at intervals of 30 min. to check beam performances. $P_{RF}$=1.6 kW, $p_{source}$=0.75 Pa, $I_{PG}$=10 A, PG-source body bias -30 V (about -0.5 A), BP-source bias -0.3 A (about -0.6 V). Plasma duration: 5 h.
- Hydrogen plasma and beam extraction for a few minutes, at time intervals from 30 min. to 10 min. (18/09/2020). The aim was to recreate the experimental conditions of pulsed negative ion source and check whether the alternation of plasma-off and plasma-on phases had a positive effect on the flux of Cs from the source [[32],[33]]. $P_{RF}$=1.6 kW, $p_{source}$=0.75 Pa, $I_{PG}$=10 A, PG-source body bias 15 V (about 0.8 A), BP-source bias 0.3 A (about 4.5 V). Test duration was 4 h, total plasma duration was 1.5 h.

During these tests the extracted beam current never exceeded 30 A/m$^2$. The line intensities of H$_\beta$, H$_\gamma$ and Cs D2 line are reported in Figure 3; not all the data are available for all the days, due to malfunctionings of the control and data acquisition system of the spectrometers. On the whole, the Cs D2 line intensity indicates a reduction of the Cs amount in the source volume during the plasma phase of the Cs cleaning tests, in particular during 15/09/2020, however the H$_\beta$/H$_\gamma$ intensity ratio did not revert its trend compared to the data of the previous days.

Table 1 reports the average values of Cs density (in units of $10^{15}$m$^{-3}$) as measured by the LAS diagnostic before, during and after the plasma phase. Variations from these values are typically within 1÷2·$10^{15}$m$^{-3}$. A large reduction of Cs density is observed in the first cleaning test during the plasma phase. The subsequent tests still show a reduction of Cs density during the plasma phase; however, the Cs density values of no-plasma phases indicate an overall reduction which develops in days, not with the action of a single cleaning test. This may indicate that the plasma is more effective in redistributing Cs inside the source, rather than in making it flow outside the source. The slow reduction of Cs may instead be related to a progressive contamination of Cs.

A final test was performed on 23/09/2020, using an argon plasma, to increase the sputtering action on the plasma on the surfaces due to the higher mass of the gas atoms. The Ar plasma duration was 3 h, at $P_{RF}$=1.0 kW, $p_{source}$=0.6 Pa, $I_{PG}$=10 A. As reported in Table 1, the Cs density before the test was lower than in the previous cleaning tests, probably because of Cs degradation occurred during the four days that passed between this test and the previous one. In the hydrogen plasma of the previous experimental session, the Cs density appeared to get lower down to a factor 2÷3 with respect to the no-plasma phase; as reported in Table 1, in the argon plasma the Cs density resulted to be up to four times higher than in the no-plasma phase, confirming the much higher sputtering capability of the more massive Ar atoms. To probe the source conditions, a hydrogen plasma was subsequently generated, at $P_{RF}$=1.5 kW, $p_{source}$=0.5 Pa, $I_{PG}$=10 A, PG-source body bias 15 V (~0.4 A), BP-source bias 0.3 A (~3.5 V). Beam extraction yielded a maximum of about 40 A/m$^2$ at $U_{ex}$=1.2 kV and $U_{tot}$=9.6 kV. During the plasma phase in hydrogen, the measured Cs density was below $10^{15}$ m$^{-3}$. However, the Cs density in the final no-plasma phase was the same as at the beginning of the day. The NIO1 source was finally disassembled to be inspected, and cleaned before continuing with further experimentation.

## III. CONCLUSIONS

The NIO1 H$^-$ source started operation with Cs evaporation in 2020, with continuous plasma phases of several hours. After reaching an extracted beam current density of 60 A/m$^2$, the beam performances progressively deteriorated because of excessive Cs evaporation into the source. The paper described the experimental evidence that was collected in this condition, and that can be useful to identify and stop overcesiation as soon as possible:

- Reduction of plasma light
- Reduction of Cs emissions during active Cs evaporation
- Reduction of the H$_\beta$/H$_\gamma$ intensity ratio, during active Cs evaporation, the other source parameters being constant.
- Impossibility to reduce the Cs density measured by LAS after a plasma session (at parity of source and PG temperatures).
- Beam perfomances (beam current, fraction of co-extracted electrons) do not change, even with no Cs evaporation (the plasma is naturally expected to worsen performances deconditioning the source).

LAS measurements proved to be fundamental to safely guide the experimentation on the source.
Attempts to clean the source with a hydrogen plasma under several conditions proved to be not sufficient with respect to the cesiation conditions reached in the source. The reduction of Cs density during the plasma phases is more likely to be attributed to a redistribution of Cs in the source rather than a net outflow of Cs from the source. LAS measurements showed that an argon plasma is more effective in sputtering Cs from the source walls, however the 3 h plasma was insufficient to remove the Cs content of the source. The test could be repeated for a longer time, to clean a Cs amount closer to the optimal conditions for NIO1.
During the cleaning test a slow reduction with time of Cs density was observed. Chemical reactions with contaminants (eg. oxygen), while potentially useful to reduce the amount of pure Cs in source, may prevent to minimize the work function at the PG surface as with pure Cs, and would be undesirable in a DEMO-like environment.

The experimentation on NIO1 will be improved in future increasing the temperature stability of the cooling system; a 100 °C heating system for the plasma grid will be introduced to improve the Cs layer conditions close to the extraction apertures [[8],[34]]. The Cs oven heating will proceed at a slower pace, from a lower $T_{res}$ range, so to find the best conditions to get closer to the extracted current density value which is ultimately required to the largest H$^-$ sources, i.e. 330 A/m$^2$ [[8],[29],[34]].